\documentclass[
twocolumn, 
superscriptaddress, 
nofootinbib, 
preprintnumbers, 
prl 
]{revtex4}

\pdfoutput=1
\usepackage{myterms}
\usepackage{mathrsfs}
\usepackage[english]{babel}
\usepackage{amsmath,amssymb,amsfonts, bm,bbm,slashed}
\usepackage{graphicx}
\usepackage{siunitx}
\usepackage[sort&compress]{natbib}
\usepackage{xcolor}
\usepackage[normalem]{ulem}
\usepackage[hypertexnames=false]{hyperref}
\usepackage[capitalize]{cleveref}
\definecolor{red}{rgb}{1.0, 0, 0}
\usepackage{enumerate}
\usepackage{epsfig, subfigure}
\usepackage{setspace}
\usepackage{booktabs, tabularx}
\usepackage{units}
\usepackage[utf8]{inputenc}
\usepackage{lineno}
\usepackage{feynmp-auto}               

\hypersetup{
        pdfnewwindow=true,   
        colorlinks=true,     
        linkcolor=blue,      
        citecolor=blue,      
        filecolor=blue,      
        urlcolor=blue        
}

\allowdisplaybreaks

\newcommand{\bra}[1]{\ensuremath{\langle #1 |}}   
\newcommand{\ket}[1]{\ensuremath{| #1 \rangle}}   
\newcommand{\ev}[1]{\ensuremath{\left\langle #1 %
                \right\rangle}} 

\newcommand{\sgn}{\text{sgn}}

\renewcommand{\vec}[1]{{\mathbf{#1}}}

\newcommand{\pvec}{\vec{p}}
\newcommand{\xvec}{\vec{x}}

\newcommand{\vw}{{v_w}}
\newcommand{\Lw}{{l_w}}
\newcommand{\gamw}{{\gamma_w}}
\newcommand{\FF}{{\mathcal{A}}}

\newcommand{\GeV}{\,\text{GeV}}
\newcommand{\TeV}{\,\text{TeV}}

\begin{document}
        

\title{
Filtered Dark Matter at a First Order Phase Transition
}
        
\author{Michael J.\ Baker}
\email{michael.baker@unimelb.edu.au}
\affiliation{School of Physics, The University of Melbourne, Victoria 3010, Australia}
\affiliation{Physik-Institut, Universit\"at Z\"urich, 8057 Z\"urich, Switzerland}

\author{Joachim Kopp}
\email{jkopp@cern.ch}
\affiliation{Theoretical Physics Department, CERN, Geneva, Switzerland}
\affiliation{PRISMA Cluster of Excellence \& Mainz Institute for Theoretical Physics, \\
Johannes Gutenberg University, Staudingerweg 7, 55099 Mainz, Germany}

\author{Andrew J. Long}
\email{andrewjlong@rice.edu}
\affiliation{Rice University, Houston, Texas 77005, USA}

\date{\today}
\preprint{ZU-TH 50/19}


\begin{abstract}
  We describe a new mechanism of dark matter production. If dark matter particles acquire mass during a first order phase transition, it is energetically unfavourable for them to enter the expanding bubbles. Instead, most of them are reflected and quickly annihilate away. The bubbles eventually merge as the phase transition completes and only the dark matter particles which have entered the bubbles survive to constitute the observed dark matter today. This mechanism can produce dark matter with masses from the TeV scale to above the PeV scale, surpassing the  Griest--Kamionkowski bound.
\end{abstract}

\maketitle


\textbf{1.~Introduction.} A wealth of observational evidence reveals that the
universe is permeated with a mysterious substance known as dark matter
(DM)~\cite{Bertone:2016nfn}.  Very little, however, is known about the particle
physics nature of DM or its origin in the early Universe.  Historically, the
favoured scenario for DM production has been thermal relic
production~\cite{Srednicki:1988ce, Gondolo:1990dk, Griest:1990kh}.  If a DM 
particle is thermalised with the Standard Model (SM) plasma in the early
Universe then the cosmological expansion, which causes the plasma to cool
adiabatically, will eventually make the DM's interactions with the SM
inefficient, driving it out of equilibrium.  Consequently, the DM
relic abundance is determined when these interactions ``freeze-out,'' typically
increasing with larger DM mass and decreasing with larger interaction
strength.  Above $m_\mathrm{DM} \sim \SI{100}{TeV}$ the required interactions
violate unitarity~\cite{Griest:1989wd,Baldes:2017gzw,Smirnov:2019ngs}.  This places an upper
bound on the mass of thermally-produced DM, known as the
Griest--Kamionkowski (GK) bound.

In this article, we propose a new mechanism for generating the DM relic
abundance.  We propose that DM freeze-out did not result from the gradual
cooling of the cosmological plasma, but instead was triggered abruptly by a
first-order cosmological phase transition (FOPT).  During the transition, DM
particles acquired a mass and low-momentum particles were ``filtered'' out of
the plasma.  We will see that DM filtration provides a viable production
mechanism, even for DM with masses above the GK bound.  

The impact of cosmological phase transitions on DM has been studied in a
variety of different contexts~\cite{Schramm:1984bt}:  a phase
transition may alter the expansion rate of the Universe during freeze-out~\cite{Kolb:1979bt,Chung:2011hv,Chung:2011it}, may inject entropy~\cite{Hambye:2018qjv,Chung:2011hv,Chung:2011it}, may alter DM stability~\cite{Baker:2016xzo,J.Baker:2018eaq,Baker:2018vos}, may alter DM
properties during freeze-in~\cite{Baker:2017zwx, Bian:2018mkl} (see also
\cite{Cohen:2008nb}), may produce DM non-thermally~\cite{Witten:1984rs,
Falkowski:2012fb, Huang:2017kzu, Bai:2018dxf}, or may produce an excess of DM 
over antimatter~\cite{Dodelson:1992rx, Shu:2006mm, Petraki:2011mv, Baldes:2017rcu,
Fornal:2017owa, Gu:2017rzz, Hall:2019rld}.  Conversely, a dark sector may trigger 
an electroweak FOPT~\cite{Gonderinger:2009jp,Carena:2011jy,Chowdhury:2011ga,Borah:2012pu,Gil:2012ya,Fairbairn:2013uta,Ahriche:2013zwa,Alanne:2014bra,Chao:2017vrq,Ghorbani:2017jls,Ghorbani:2019itr}.  
Freeze-out during a second order phase transition
has been studied in Ref.~\cite{Dimopoulos:1990ai,Heurtier:2019beu},
and Ref.~\cite{Dvali:1997sa} used domain walls to ``sweep away''
over-abundant magnetic monopoles.

Ref.~\cite{Hambye:2018qjv} recently studied a model where DM acquires mass during a strongly supercooled FOPT and its relic abundance was suppressed by the associated entropy injection.  By contrast, our interest is in the dynamical interaction of DM particles with bubble walls and its impact on the relic abundance.


\begin{figure}
  \begin{center}
    \includegraphics[width=0.95\columnwidth]{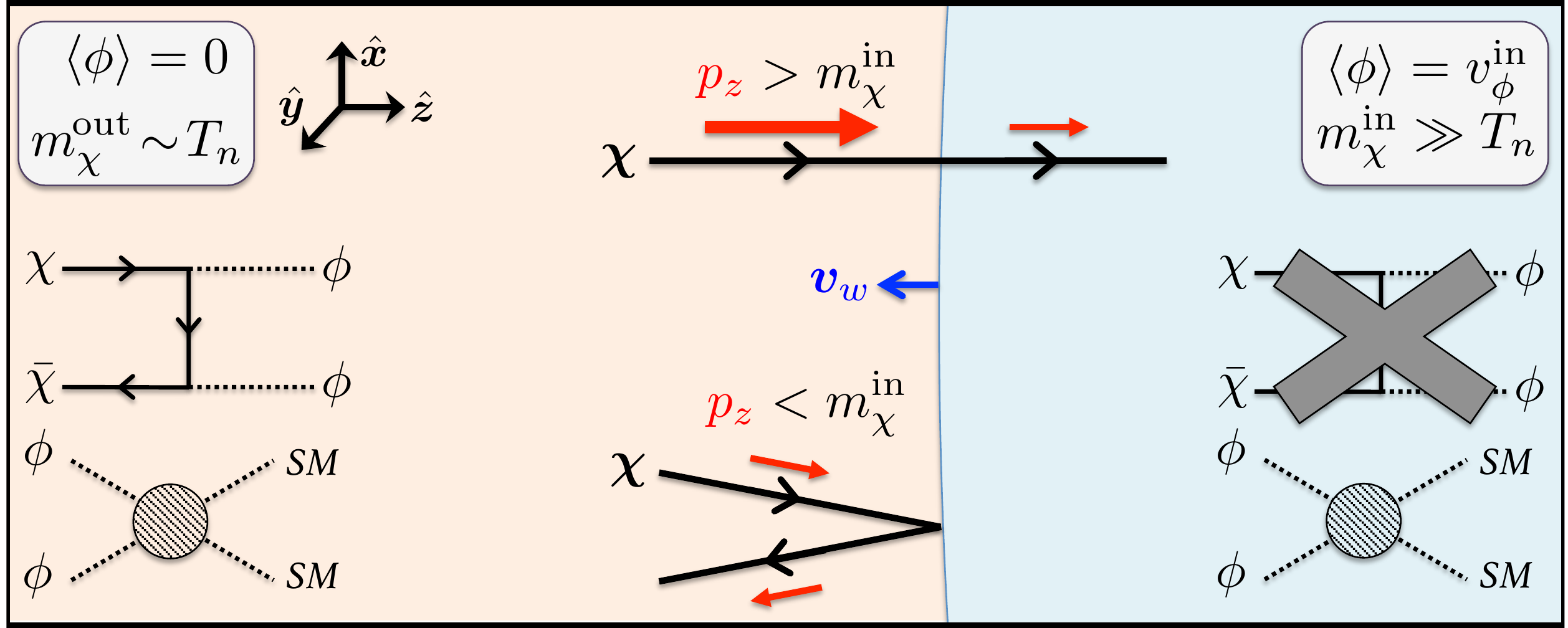} 
  \end{center}
  \caption{\label{fig:cartoon} ``Filtered DM'':
    only DM particles $\chi$ with kinetic energy $E \gtrsim m_\chi^\mathrm{in}$
    can penetrate the bubble; slower particles are reflected.  In
    front of the bubble wall (pink region), DM is kept in thermal
    equilibrium through $\chi \bar{\chi} \leftrightarrow \phi \phi$, but this
    reaction is put abruptly out of equilibrium at the wall where $\chi$
    obtains a mass (blue region).  
    The field $\phi$ remains in equilibrium throughout. 
    }
\end{figure}

\textbf{2.~The mechanism.}
Our proposed mechanism for DM filtration during a FOPT is illustrated in
\cref{fig:cartoon}.  DM particles $\chi$ initially have a small mass $m_\chi^\text{out} \sim T$ 
and are in thermal equilibrium with SM particles and a new scalar particle
$\phi$.  We imagine that $\phi$ undergoes the FOPT at temperature
$T_n$:  its thermal expectation value is initially vanishing, $\ev{\phi} = 0$,
but jumps to a non-zero value, $\ev{\phi} = v_\phi^\mathrm{in}$, during the
FOPT.  FOPTs proceed through the nucleation and growth of bubbles of
the new $\ev{\phi} = v_\phi^\mathrm{in}$ phase ~\cite{Linde:1981zj}.  These
bubbles expand and merge until the whole universe has transitioned.  
At the interface of the old and new phase there is a bubble wall where
$\ev{\phi}$ smoothly transitions from zero to $v_\phi^\mathrm{in}$. 

We assume that $\ev{\phi} \neq 0$ generates a large mass for the DM particles, 
so light DM particles
become heavy as they cross the wall into the bubble.  Energy conservation
implies that a DM particle can only penetrate the bubble wall if its kinetic
energy $E \gtrsim m_\chi^\mathrm{in}$.  Lower momentum modes are
reflected by the advancing bubble walls.  If $m_\chi^\mathrm{in} \gg T$, then
only an exponentially small fraction of the DM particles will have enough
kinetic energy to enter the bubbles.  As DM particles enter the bubble, their
interactions are put abruptly out of equilibrium, preventing their
annihilation. DM particles outside the bubble in contrast will continue to
interact efficiently, so that the reflected particles 
quickly annihilate away into the thermal bath.  Once the broken phase permeates
the whole universe, only the particles that have entered the bubbles remain and
constitute the DM observed today.


\textbf{3.~A toy model.}
To derive quantitative results we introduce a toy model, which is 
a viable theory of DM in its own right.  We augment the SM by a
gauge-singlet real scalar field $\phi(x)$ and a singlet Dirac spinor field $\chi(x)$.  
The Lagrangian defining this theory contains the 
terms
\begin{align}
  \Lscr &\supset - V(\phi) 
                 - y_\chi \phi \bar{\chi} \chi
                 - \beta \, \phi^2 H^\dagger H
        \com
  \label{eq:model}
\end{align}
where $V(\phi)$ is the scalar potential, $y_\chi$ is a real Yukawa coupling,
$\beta$ is a real Higgs portal coupling, and $H(x)$ is the SM Higgs field.  We
do not assume any particular form for $V(\phi)$, only that it gives $\phi$ a
mass $m_\phi$ and causes a FOPT in which $\phi$ acquires a non-zero vacuum
expectation value $\bra{0} \phi \ket{0} = v_\phi$.  Typically
$v_\phi^\mathrm{in} \lesssim v_\phi$.  For simplicity we assume that the mass
of $\phi$ does not change appreciably during the FOPT.
Note that $\chi$ enjoys a global $U(1)$ symmetry that ensures its stability.  

Before the FOPT the Yukawa interaction leads to a thermal mass for $\chi$,
$m_\chi^\text{out} = y_\chi T/4$, while afterwards it also induces a larger mass 
$m_\chi^\mathrm{in} \sim y_\chi v_\phi^\mathrm{in}$ 
(we are interested in regimes where $y_\chi v_\phi^\mathrm{in}  \gg T_n \sim m_\phi$).  
The Yukawa interaction allows
$\chi$ to annihilate, chiefly via $\chi \bar{\chi} \to \phi \phi$, 
while the thermal mass typically forbids the 
process $\chi \bar{\chi} \leftrightarrow \phi$.  
In the following we retain this condition  but otherwise approximate $m_\chi^\text{out} = 0$.  
We treat $v_\phi^\mathrm{in} / T_n$ as a free parameter, since we do not specify
the form of $V(\phi)$, but we remark that large order parameters may arise from
nearly conformal potentials~\cite{Creminelli:2001th, Nardini:2007me,
Konstandin:2011dr} or models with heavy fermions (such as $\chi$
here)~\cite{Carena:2004ha, Angelescu:2018dkk}.  

The Higgs portal interaction~\cite{Silveira:1985rk, Burgess:2000yq,
Patt:2006fw} in \cref{eq:model} allows the hidden sector to communicate with
the SM, through reactions such as $\phi \phi \leftrightarrow H^\dagger H$ if
$m_\phi$ is above the Higgs mass, $m_h$, and $\phi \phi \leftrightarrow f \bar{f}$
if not.  We ensure that $\beta$ is large enough to thermalise $\phi$ and the SM
at a common temperature $T_n$ during the FOPT.  At later times the
Higgs portal interaction allows $\phi$ particles to decay to SM particles.
If $m_\phi < m_h/2 \simeq 62.5 \GeV$, the Higgs portal coupling is constrained
to be $\beta \lesssim 0.007 (1 -
4m_\phi^2/m_h^2)^{-1/4}$~\cite{Sirunyan:2018owy} whereas $\beta$ is almost
entirely unconstrained if $m_\phi > m_h/2$.

A relatively large $m_\chi^\mathrm{in}$ ensures that $\chi \bar{\chi}
\leftrightarrow \phi\phi$ is out of equilibrium inside the bubble.  If this
were not the case,  $\chi$ would remain in thermal equilibrium through the
FOPT and its relic abundance would later be determined by standard
thermal freeze-out.  We therefore require the thermally-averaged annihilation
rate $\Gamma$ to be smaller than the cosmological expansion rate $H$ inside the
bubble.  This leads to the condition
\begin{align}
  \frac{m_\chi^\mathrm{in}}{T_n}
    &\gtrsim 24 - \log \frac{T_n}{\si{TeV}} - \frac{3}{2} \log\frac{m_\chi^\mathrm{in}/T_n}{24}
                + 4 \log y_\chi \,,
  \label{eq:OOE-condition}
\end{align}
where we have used $H = (\pi/\sqrt{90}) \, \sqrt{g_\ast} \, T_n^2 / \Mpl$ and
$\Gamma = \ev{\sigma v} n_\chi^\mathrm{in,eq}$, with the thermally
averaged annihilation cross section
$\ev{\sigma v} \simeq (9 \, y_\chi^4 \, T_n) / (64\pi \,
(m_\chi^\mathrm{in})^3)$ \cite{Gondolo:1990dk} and the would-be equilibrium abundance
$n_\chi^\mathrm{in,eq} = g_\chi (m_\chi^\mathrm{in} T_n
/ 2\pi)^{3/2} \, e^{-m_\chi^\mathrm{in}/T_n}$.
$g_\chi = 2$ counts the spin states, $g_{\ast} \simeq 100$ is the
effective number of relativistic species, and $\Mpl \simeq \SI{2.43e18}{GeV}$
is the reduced Planck mass.
Since $m_\chi^\mathrm{in} = y_\chi \, v_\phi^\mathrm{in}$, \cref{eq:OOE-condition} allows $y_\chi = O(1)$ and $v_\phi^\mathrm{in} / T_n = O(10)$; smaller $y_\chi$ needs larger $v_\phi^\mathrm{in} / T_n$.


\textbf{4.~Analytic estimates.}
We first estimate the DM relic abundance by employing a simplified description
of the FOPT dynamics,  treating the $\chi$ particles as they
interact with the wall as if they were free particles. In other words, we
assume that the thickness of the bubble wall, $\Lw$ is much smaller than the DM
interaction length $l_\text{int}$.  Due to energy conservation, the mass
increase of $\chi$ particles crossing the wall implies that only high-momentum
particles can enter the bubble, while low-momentum ones will be reflected.
After a distance $l_\text{int}$ these reflected particles will be absorbed back
into the thermal bath, so low-momentum $\chi$ particles are filtered
out of the plasma by the wall.  Both reflected and penetrating particles
transfer momentum to the bubble wall, leading to friction that limits the speed
at which the wall advances, $\vw$~\cite{Moore:1995ua,Moore:1995si}.  

Using energy and transverse momentum conservation, we find that a massless $\chi$ 
particle that's incident on the wall with momentum $\pvec = (p_x, p_y, p_z)$ (in the
plasma's rest frame) will only have sufficient energy to enter the bubble
if $\gamw ( p_z + \vw \, |\pvec| ) > m_\chi^\mathrm{in}$~\cite{Bodeker:2009qy}, 
where $\gamw = 1 / \sqrt{1 - \vw^2}$ is the wall's Lorentz factor and we have 
assumed the wall moves in the negative $z$ direction.  Once such a particle 
enters the bubble it slows down to travel with a speed 
$v_\chi^\mathrm{in} = [\, |\pvec|^2 - (m_\chi^\mathrm{in})^2 \,]^{1/2} / m_\chi^\mathrm{in}$.  
We will be interested in non-relativistic walls, $\vw \lesssim 0.1$ because of 
the aforementioned friction effect.  Moreover, if the wall moves relativistically, 
most $\chi$ particles enter the bubble. 

If a thermal flux of $\chi$ particles is incident on the wall, the number
density $n_\chi^\mathrm{in}$ of $\chi$ particles that have entered the bubble is
\begin{align}
  n_\chi^\mathrm{in} = n_{\bar\chi}^\mathrm{in}
    &= g_\chi \int \!\! \frac{\ud^3 \pvec}{(2\pi)^3} \,
                   \frac{\Theta( p_z + \vw \, |\pvec|
                                  - m_\chi^\mathrm{in} / \gamw)}{e^{|\pvec|/T_n} + 1} \, \frac{1}{v_\chi^\mathrm{in}}
                                                                    \notag\\
    &\approx \frac{g_\chi (m_\chi^\mathrm{in} T_n)^{3/2}}{4(2\pi)^{3/2}} \, e^{-m_\chi^\mathrm{in} / T_n} 
     =       \frac{1}{4} n_\chi^\mathrm{in,eq}\,,
  \label{eq:n_chi_in}
\end{align}
where the step function $\Theta$ enforces the kinematic condition above,
$n_\chi^\mathrm{in,eq}$ was defined below \cref{eq:OOE-condition}, and
$1/v_\chi^\mathrm{in}$ accounts for the reduced speed of particles inside the
bubble.  The Boltzmann-like exponential factor is crucial in suppressing the
abundance of DM inside the bubbles and therefore in setting the relic
abundance.  In front of the bubble wall, reflected DM annihilates 
$\chi \bar{\chi} \to \phi \phi$, and $\phi$ remains in equilibrium.  
The associated entropy transfer and heating are negligible if $g_\ast = O(100)$.  

Since $\chi \bar{\chi} \leftrightarrow \phi \phi$ is out of equilibrium inside
the bubble, the $\chi$ and $\bar{\chi}$ particles that enter during the phase
transition will survive until today, where they constitute the relic population
of DM.  The corresponding relic abundance $\Omega_\mathrm{DM}$ is calculated by
scaling $n_\chi^\mathrm{in} + n_{\bar{\chi}}^\mathrm{in}$ with the entropy
density $s = (2\pi^2/45) g_{\ast S} T^3$, where $g_{\ast S} = g_\ast$ at $T_n$
and $g_{\ast S} = g_{\ast S 0} \equiv 3.9$ today (see also Ref.~\cite{Nakai:2017qos}).  
After normalizing to the
critical density $\rho_c = 3 H_0^2 \Mpl^2$, we obtain
\begin{align}
  \Omega_\mathrm{DM} h^2
    &\simeq \frac{m_\chi (n_\chi^\mathrm{in} + n_{\bar{\chi}}^\mathrm{in})}{3 \Mpl^2 (H_0/h)^2}
            \frac{g_{\ast S 0} T_0^3}{g_{\ast S} T_n^3} \notag\\
    &\simeq 
    0.17 \, 
    \bigg( \frac{T_n}{\text{TeV}} \bigg)
    \bigg( \frac{m_\chi^\mathrm{in} / T_n}{30} \bigg)^{5/2}
    \frac{e^{- m_\chi^\mathrm{in} / T_n}}{e^{-30}} \,,
  \label{eq:Omega_chi}
\end{align}
where $H_0 = 100 \, h \km / \text{sec} / \text{Mpc}$ is the Hubble constant and
$T_0 \simeq \SI{0.235}{meV}$ is the temperature of the cosmic microwave background
today.  In obtaining this estimate, we have neglected the heating of the SM
bath by the annihilation of the reflected $\chi$ particles in front of wall and
by the eventual decay of $\phi$. This is justified because the number of SM
degrees of freedom at $T_n \gtrsim \text{GeV}$ is much larger than the number
of dark sector degrees of freedom.  The observed DM relic abundance,
$\Omega_\mathrm{DM}^\mathrm{obs.} h^2 \simeq 0.12$~\cite{Aghanim:2018eyx},
is obtained if the DM mass
increases to $m_\chi^\mathrm{in} \sim 30\, T_n$ inside the bubble for $T_n \sim
1\TeV$.  At higher (lower) phase transition temperatures, the required
$m_\chi^\mathrm{in} / T_n$ becomes larger (smaller), but only logarithmically
due to the exponential suppression.  Comparing \cref{eq:Omega_chi}
against the standard thermal freeze-out calculation, we note that our predicted
relic abundance only depends on the DM's interaction strength, $y_\chi$,
through $m_\chi^\mathrm{in} / T_n = y_\chi v_\chi^\mathrm{in} / T_n$, and
consequently there is a not a one-to-one mapping from the parameters that set
the relic abundance to the parameters probed, for instance, by direct detection
experiments.  

Inside the bubbles, DM could be produced from freeze-in~\cite{McDonald:2001vt,Hall:2009bx,Hambye:2018qjv}, but for typical parameters $\Omega_\chi h^2 \sim 10^{-6} (y_\chi / 2)^4 e^{-2(m_\chi^\mathrm{in}/T_n-32)}$, making this population negligible.


\textbf{5.~Numerical solution of Boltzmann's equation.}  
To obtain a more accurate estimate of the relic abundance, we numerically solve
the Boltzmann equations describing the $\chi$ particles near the bubble wall (see Supplemental Material).  

Since the scattering and diffusion length scales are small compared to the
curvature scale of a typical bubble, we assume that the bubble wall is planar,
and take the wall to be perpendicular to the $z$-axis.  Since the wall
experiences a significant drag force from the the scattering of $\chi$
particles, we assume a constant non-relativistic (terminal) wall speed, $\vw$.
We choose $\vw = 0.01$ but have checked that the final relic abundance is not
strongly dependent on its precise value.  We approximate the mass profile of DM
particles across the wall with a smoothed step-function, $m_\chi(z) =
\tfrac{1}{2} m_\chi^\mathrm{in} [ 1 + \tanh (3z / \Lw )]$.  Here and in the
remainder of the article we work in the wall's rest frame.  We use a wall
thickness $\Lw = 1 / (4 T_n)$, but find that the final relic abundance does not
depend strongly on the precise value.   

Let $f_a(t,\xvec,\pvec)$ be the phase space distribution functions for $a =
\chi$, $\bar{\chi}$, and $\phi$ particles.  We assume that the conserved
$\chi$--$\bar{\chi}$ asymmetry is vanishing, thus $f_{\bar{\chi}} = f_\chi$, and
that $\phi$ remains in equilibrium throughout the FOPT: $f_\phi =
f_\phi^\mathrm{eq}$ is the Bose-Einstein distribution.  This is justified
provided that $\phi$
depletion is fast enough to keep up with $\phi$ production.  
Far in front of the wall ($z \to - \infty$), $f_\chi =
f_\chi^\mathrm{eq}$ follows the Fermi-Dirac distribution.  We adopt the
ansatz,  
\begin{align}
  & f_\chi(z,\pvec) = \FF(z,p_z) \times f_\chi^\mathrm{eq}(z,\pvec)\,,
  \label{eq:fchi}
\end{align}
motivated in the Supplemental Material. The
distribution $f_\chi$ in the vicinity of the bubble wall can then be
described by the Boltzmann equation
\begin{widetext}
\begin{align}
  &\Bigg[\Bigg(
     \frac{p_z}{m_\chi} \frac{\partial}{\partial z}
   - \bigg( \frac{\partial m_\chi}{\partial z} \bigg) \frac{\partial}{\partial p_z}
   - \bigg( \frac{\partial m_\chi}{\partial z} \bigg) \frac{\vw}{T_n}
   \Bigg) \FF(z, p_z) \Bigg]
   \frac{g_\chi m_\chi T_n}{2\pi}
    \exp\Biggl[ \frac{\vw p_z - \sqrt{m_\chi^2 + (p_z)^2}}{T_n} \Biggr]
  = g_\chi \int \! \frac{\ud p_x \, \ud p_y}{(2\pi)^2} \, \mathbf{C}[f_\chi] \,.
  \label{eq:Boltzmann_eqn}
\end{align}
\end{widetext}
The right hand side includes the collision terms for the processes $\chi
\bar{\chi} \to \phi \phi$, $\chi \phi \to \chi \phi$, $\chi \chi \to \chi \chi$
and $\chi \bar{\chi} \to\chi \bar{\chi}$.  
Note that we have integrated over $p_x$ and $p_y$.  
Integrating over $p_z$ will then yield the number density at a position $z$.

\begin{figure}
  \begin{center}
   \includegraphics[width=0.85\columnwidth]{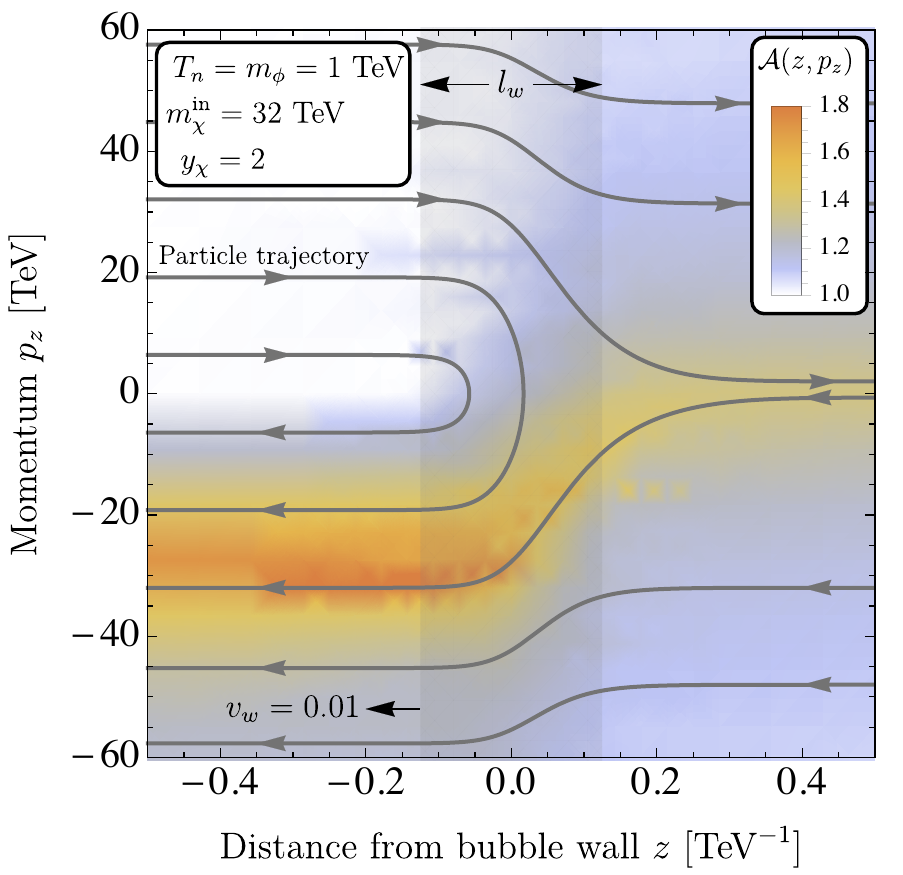} 
  \end{center}
  \caption{
    The enhancement factor $\FF(z, p_z)$ in the neighbourhood of
    the bubble wall (opaque vertical band).  Contours with arrows indicate
    possible particle trajectories in this two-dimensional phase space.  For
  the chosen parameter values, we recover the observed relic abundance. 
  }
  \label{fig:soln}
\end{figure}

We are interested in solutions of \cref{eq:Boltzmann_eqn} that obey the
boundary conditions 
\begin{align}
  \lim_{\substack{z \to -\infty \\ p_z > 0}} \FF \to 1 
  \hspace{11pt} \text{and} \hspace{11pt}
  \lim_{z \to \infty} \FF(p_z) = \lim_{z \to \infty} \FF(-p_z) \,.
  \label{eq:boundary_condit}
\end{align}
The first condition enforces an equilibrium phase space distribution for
particles that have not yet interacted with the bubble wall, while the second
condition is based on the assumption that at a large positive $z$ the other
side of the bubble is advancing with similar dynamics.  We solve
\cref{eq:Boltzmann_eqn,eq:boundary_condit} numerically using the method of
characteristics, where the 2-dimensional partial differential equation is
re-written as an infinite set of uncoupled ordinary differential equations.
Each equation corresponds to a possible particle trajectory in the
two-dimensional phase space spanned by $z$ and $p_z$, in the absence of
collisions.  A typical solution for $\FF(z, p_z)$ is shown in \cref{fig:soln},
along with some of the aforementioned particle trajectories.  Particles
incident on the wall begin in equilibrium (upper-left quadrant), so $\FF(z,
p_z) \approx 1$.  
Those that started with a momentum larger than 
$m_\chi^\mathrm{in}$ enter the bubble (upper-right), with
$\FF \approx 1.2$. 
That is, with an abundance only slightly larger than the
strongly Boltzmann-suppressed $f_\chi^\mathrm{eq}(z, \pvec)$.  Particles
that started with a momentum lower than $m_\chi^\mathrm{in}$ are reflected by the
wall (mid-left).  
Particles that come from $z \to \infty$ (lower-right) receive a boost in momentum as
they leave the bubble.  
These boosted particles and the reflected particles lead to an overdensity, 
which annihilates into the thermal bath as the particles travel away 
from the wall (bottom-left).

\begin{figure}
  \begin{center}
    \includegraphics[width=0.95\columnwidth]{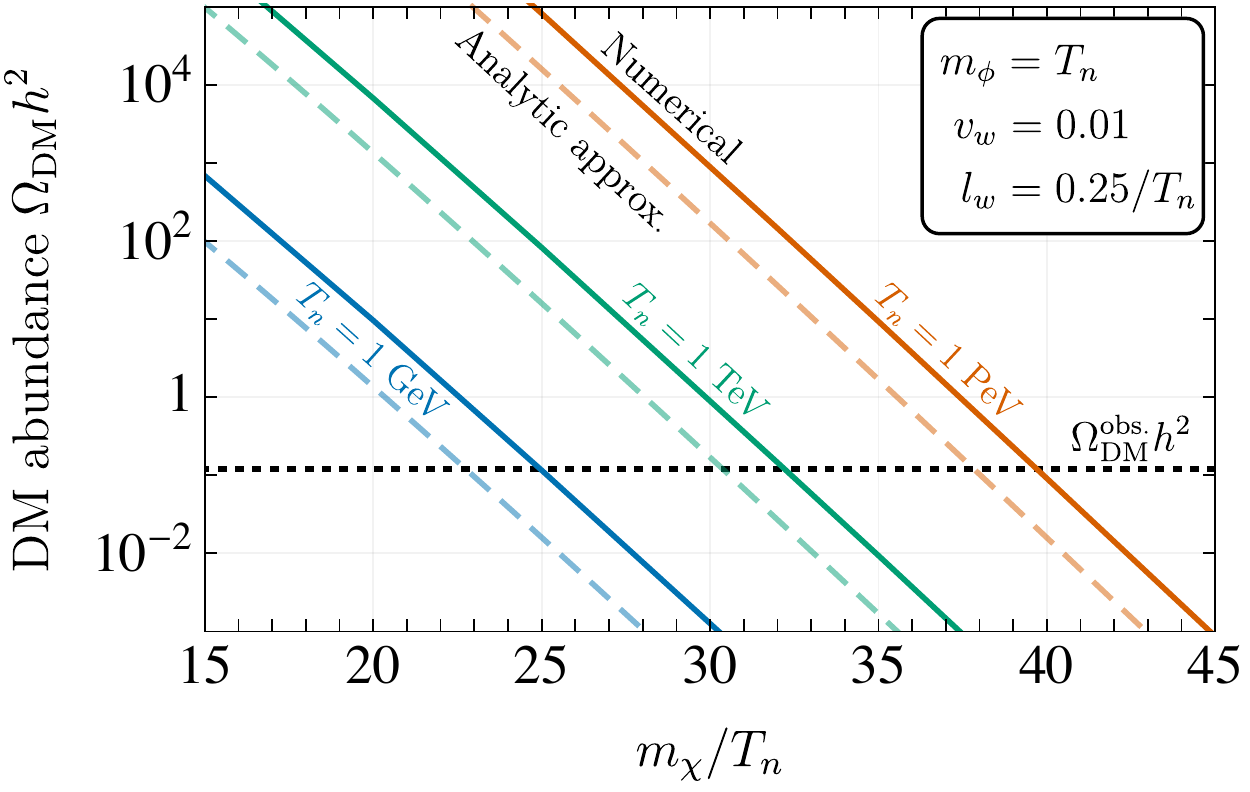} 
  \end{center}
  \caption{ The  DM relic abundance as a function of the FOPT's
    temperature $T_n$ and the $\chi$ particle's mass $m_\chi$, where we assume
    $m_\chi \approx m_\chi^\mathrm{in}$.  The solid lines are calculated by
    numerically solving Boltzmann's equation while the dashed lines show the
    analytic approximation, \cref{eq:Omega_chi}.}
\label{fig:omh2}
\end{figure}

We then integrate over $p_z$ deep inside the bubble to find the resulting DM
relic abundance, and present our results in \cref{fig:omh2}.  We assume $m_\chi
\approx m_\chi^\mathrm{in}$, implying a negligible change in the $\chi$
particle's mass between the FOPT and today.  The observed relic
abundance is obtained for $m_\chi/T_n \approx$ (25, 32, 40) and $T_n =$
(\SI{1}{GeV}, \SI{1}{TeV}, \SI{1}{PeV}), respectively.  These parameters are
consistent with the out-of-equilibrium condition \cref{eq:OOE-condition}
provided that $y_\chi < (0.2, \sqrt{4\pi}, \sqrt{4\pi})$,
respectively.  The exponential sensitivity to $m_\chi/T_n \approx
m_\chi^\mathrm{in}/T_n$ is clearly visible.  Comparing the numerical result
with the analytical estimate from \cref{eq:Omega_chi}, we find good agreement
of the parametric dependences on $T_n$ and $m_\chi^\mathrm{in}/T_n$,
and the overall amplitude differs by a factor of $\sim 5$.


\begin{figure}
  \begin{center}
    \includegraphics[width=0.95\columnwidth]{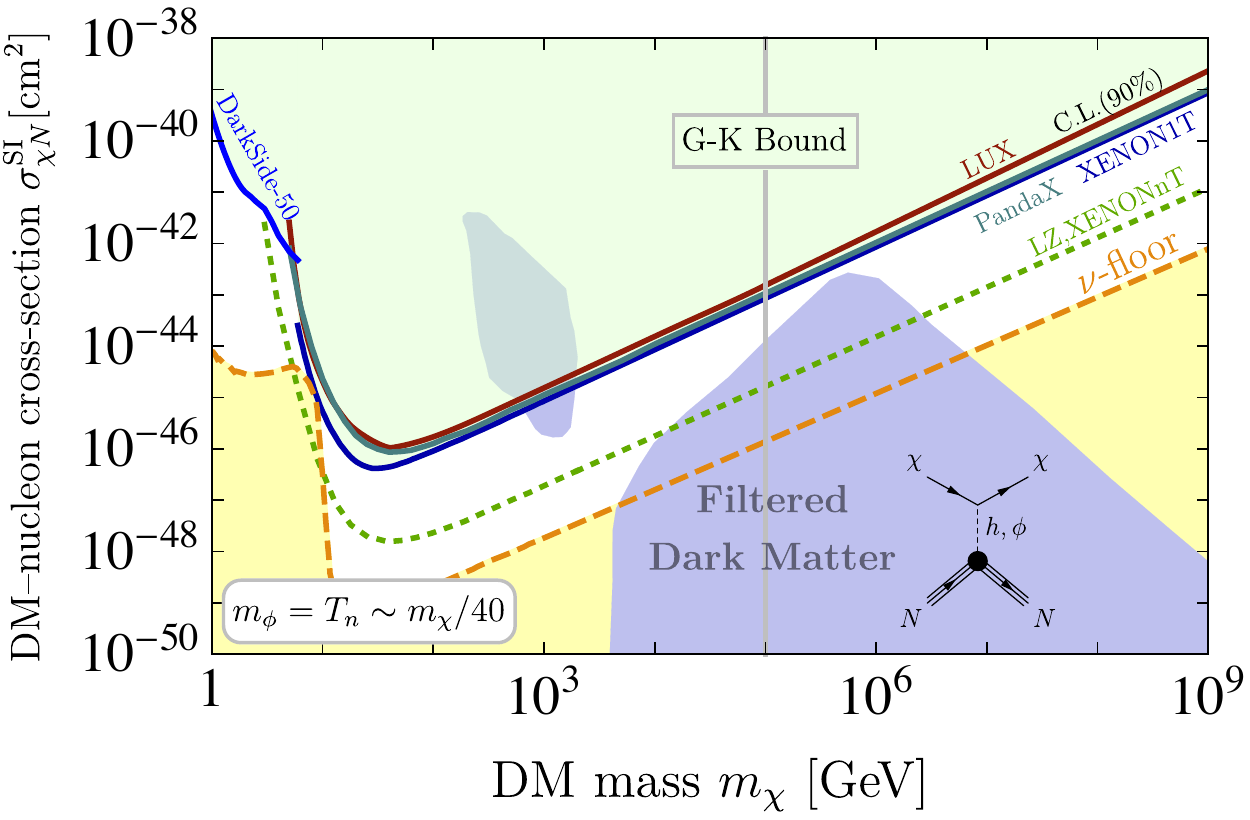} 
  \end{center}
  \caption{The predicted spin-independent DM--nucleon scattering cross-section
    (purple shaded region) in comparison with various experimental exclusions
    limits (green shaded)~\cite{Akerib:2016vxi, Cui:2017nnn, Aprile:2018dbl,
    Agnes:2018ves}, projected sensitivities of future experiments (green
    dashed) \cite{Akerib:2018lyp}, and the neutrino floor (yellow shaded).
    Note that viable models of filtered DM are obtained even at 
    DM masses above the Griest--Kamionkowski bound,
    $m_\chi \sim 100 \TeV$.
  }
  \label{fig:dd-cs}
\end{figure}

\textbf{6.~Current and future probes.}
Filtered DM is amenable to many of
the same tests as thermal relic (WIMP) DM.  Direct detection of $\chi$
particles is mediated, in this toy model, 
via exchange of $\phi$ particles and Higgs bosons
($h$) so the rate is suppressed by the tiny $\phi$--$h$ mixing~\cite{Matsumoto:2018acr}.  In
\cref{fig:dd-cs} the purple region shows the range of spin-independent
$\chi$--nucleon scattering cross-sections $\sigma_{\chi N}^\mathrm{SI}$.
We impose the conditions that $\Omega_\chi = \Omega_\mathrm{DM}^\mathrm{obs.}$, 
that couplings remain perturbative ($y_\chi, \beta <
\sqrt{4\pi}$), that $\chi$ is in equilibrium outside the bubble 
and out of equilibrium inside the
bubble, \cref{eq:OOE-condition}, and that $\phi$ is in equilibrium 
throughout the FOPT.  
At $m_\chi \ll \SI{100}{GeV}$, the dark sector no longer stays
in equilibrium outside the bubble because $\phi\phi$ annihilation
is suppressed by the Higgs mass and small SM Yukawa couplings.
Around masses of several \si{TeV}, the value of $\beta$ required to
keep $\phi$ in equilibrium grows, making it impossible to obtain the
correct Higgs mass from the scalar mass matrix. At even larger $m_\chi$,
this problem disappears as new $\phi$ annihilation channels open up.
We see that there is a large region of viable parameter space
at masses above the Griest--Kamionkowski bound~\cite{Griest:1989wd,Baldes:2017gzw}.

At current and future collider experiments, filtered DM can be tested
through precision measurements of the Higgs boson's couplings to other SM
particles~\cite{Fujii:2017vwa, CEPCStudyGroup:2018ghi, Abada:2019lih,
Cepeda:2019klc}.  These measurements already constrain the $\phi$--$h$ mixing
for sub-TeV masses~\cite{Khachatryan:2016vau,Carena:2018vpt}.  

Annihilations of $\chi$ and $\bar{\chi}$ to SM particles in the Milky Way's DM
halo provide another avenue to indirectly detect filtered DM. Decays of the annihilation products may be a source of PeV-scale neutrinos.
Detection prospects are however hampered by $p$-wave suppressed annihilation
cross-sections.  

The FOPT bubble dynamics produce a stochastic background of gravitational
waves~\cite{Kamionkowski:1993fg}.  The frequency of this 
radiation is tied to the DM mass scale.  However, we
expect the signal strength to be suppressed by the small bubble wall speed and
a dedicated analysis is required to determine if this signal is within reach of
next-generation gravitational wave telescopes, e.g., 
LISA~\cite{Caprini:2019egz}.


\begin{acknowledgments}
\textit{
  Acknowledgments.  The authors would like to thank Andrea Thamm for comments
  on the manuscript and Christopher Tunnell for discussions of direct detection prospects.  
  The authors would also like to express a special thanks
  to the Mainz Institute for Theoretical Physics (MITP) for its hospitality and
  support during key parts of the collaboration.
  M.J.B.\ was supported by the Australian Research Council and by the Swiss
  National Science Foundation (SNF) under contract 200021-159720.  J.K.\ has been
  partially supported by the European Research Council (ERC) under the European
  Union's Horizon 2020 research and innovation program (grant agreement No.\
  637506, ``$\nu$Directions'') and by the German Research Foundation (DFG) under
  grant No.\ KO~4820/4-1.  A.J.L.\ was supported in part by the US Department of
  Energy under grant DE-SC0007859.  
}
\end{acknowledgments}


\bibliographystyle{JHEP}
\bibliography{./refs}


\clearpage
\newpage
\maketitle
\onecolumngrid
\begin{center}
\textbf{\large Filtered Dark Matter at a First Order Phase Transition} \\ 
\vspace{0.05in}
{\it \large Supplementary Material}\\ 
\vspace{0.05in}
{Michael J.~Baker, \ Joachim Kopp, and Andrew J.~Long}
\end{center}
\onecolumngrid

\setcounter{equation}{0}
\setcounter{figure}{0}
\setcounter{table}{0}
\setcounter{section}{0}
\setcounter{page}{1}
\makeatletter
\renewcommand{\theequation}{S\arabic{equation}}
\renewcommand{\thefigure}{S\arabic{figure}}


\hfill \parbox{14.8cm}{
\begin{centering}
  In this supplemental material to our Letter, ``Filtered Dark Matter at a First Order Phase Transition'', we provide details of our  
  derivation and solution of the relevant Boltzmann equations.  We also discuss the annihilation of dark matter particles that are 
  reflected by the bubble walls.  
  \end{centering}
  } \hfill \phantom{.}


\section{Derivation of the Boltzmann equations}
\label{sec:derivation-of-boltzmann}

Here we present a derivation and further discussion of the Boltzmann equation,
Eq. (6) in the main text.  We start from the general
Boltzmann equation for the phase space distribution of $\chi$ particles, $f_\chi$,
\begin{align}
  \mathbf{L}[f_\chi] &= \mathbf{C}[f_\chi] \,.
  \label{eq:boltzmann}
\end{align}
Here, $\mathbf{L}$ is the Liouville operator, which describes the evolution of
$f_\chi$ in the absence of particle scattering, and $\mathbf{C}$ is a collision
term, which accounts for any particle number changing processes as well as
elastic scattering.

Before discussing these terms in more detail, we motivate the ansatz that we
will use, Eq. (5) in the main text. We start with the Maxwell-Boltzmann
approximation for the equilibrium distribution function, which in the plasma
frame is
\begin{align}
f_\chi^\mathrm{eq}
          \approx  \exp \left(\frac{\mu}{T}\right) \exp\left(-\frac{E^p}{T}\right)\,,
\end{align}
where $\mu$ is the chemical potential, the energy $E^p = \sqrt{m(z^p, t^p)^2 +
(\vec{p}^p)^2}$, and $z^p$, $t^p$ and $\vec{p}^p$ are the position
perpendicular to the wall, the time coordinate and the 3-momentum in the plasma
frame.  The Maxwell-Boltzmann approximation is very good in the
non-relativistic regime and gives an error of $\approx$ 5\% for both fermions
and bosons in the ultra-relativistic regime where the mass is negligible and
the energy $\approx 3T$.  Since in the rest frame of the wall the mass of
$\chi$ is a function of the position perpendicular to the wall, $z^w$,  alone,
it is natural to work in that frame.  The energy of a particle in the wall
frame $E^w$ is related to the energy in the plasma frame by
\begin{align}
  E^p &= \gamw \left(E^w - \vw p_z^w \right)\,,
\end{align}
where ${\bm v}_w = - \vw \, \hat{\bm z}$ is the bubble wall velocity (in
natural units) and $\gamw = 1 / \sqrt{1 - \vw^2}$ is the corresponding Lorentz
factor.  We take the wall to be moving in the negative $z$ direction, so the
wall speed $\vw$ is positive.  We expect interactions with the wall to cause
deviations from equilibrium.  In order to describe these deviations in front of
the wall it is essential to retain the dependence on $z^w$ and $p_z^w$ (the
momentum perpendicular to the wall in the wall frame).  For simplicity, we
will, however, assume that the pre-factor $\FF$ which describes the deviation
from equilibrium and also absorbs the chemical potential term,
does not depend on $p_x$ or $p_y$. This motivates the ansatz
\begin{align}
  f_\chi &= \FF(z^w,p_z^w) \exp\left(-\frac{E^p}{T}\right)\,.
  \label{eq:ansatz}
\end{align}
While this is related to the fluid ansatz~\cite{Moore:1995si}, we take the
temperature to be homogeneous, but introduce dependence on $p_z^w$.  Although
$\chi$ is not in equilibrium inside the bubble, we see from Eq. (3) in the main text 
that we expect an order one deviation from equilibrium inside the bubble, so
$\FF(z^w,p^w_z)$ should not deviate dramatically from one.

\subsection{The Liouville Operator}

The Liouville operator is the total time derivative of the phase space 
distribution function $f = f(t, \xvec(t), \pvec(t))$. In the wall frame,
it is given by
\begin{align}
  \mathbf{L}[f_\chi] &= \frac{df_\chi}{dt^w}
                      = \frac{\partial f_\chi}{\partial t^w}
                      + \frac{\partial \xvec^w}{\partial t^w}
                                     \frac{\partial f_\chi}{\partial \xvec^w}
                      + \frac{\partial \pvec^w}{\partial t^w}
                                     \frac{\partial f_\chi}{\partial \pvec^w} \,.
  \label{eq:LiouvilleTotDeriv}
\end{align}
We assume that the system has reached a steady-state 
($\partial f_\chi / \partial t^w = 0$) in the wall frame.  
Since the system is translation invariant in $x$ and $y$, 
we can simplify $\mathbf{L}[f_\chi]$ considerably:
\begin{align}
  \mathbf{L}[f_\chi] &= \frac{p_z^w}{E^w} \frac{\partial f_\chi}{\partial z^w}
                      + \frac{\partial p_z^w}{\partial t^w} \frac{\partial f_\chi}{\partial p_z^w} \,,
\end{align}
where we have written the velocity $\partial z^w / \partial t^w$ as $p_z^w / E^w$.  
The factor $\partial p_z^w / \partial t^w$ can be interpreted as the semi-classical 
force acting on the particle as it traverses the wall.  Since 
$E^2 = \pvec_\perp^2 + p_z^2 + m^2$ and since the particle energy and the transverse
momentum $\pvec_\perp$ are conserved in the wall frame we can write
\begin{align}
  \frac{\partial p_z^w}{\partial t}
  &=  \frac{\partial (\sgn[p_z^w]\sqrt{(E^w)^2 - p_\perp^2 - m(z^w)^2})}{\partial t}\notag\\
  &= -\frac{m(z^w)}{p_z^w} \frac{\partial z^w}{\partial t^w}
      \frac{\partial m(z^w)}{\partial z^w} \notag\\
  &= -\frac{m(z^w)}{E^w} \frac{\partial m(z^w)}{\partial z^w} \,.
\end{align}
We then have
\begin{align}
  \mathbf{L}[f_\chi] &= \frac{p_z^w}{E^w} \frac{\partial f_\chi}{\partial z^w}
                      - \frac{m(z^w)}{E^w} \frac{\partial m}{\partial z^w}
                        \frac{\partial f_\chi}{\partial p_z^w}\,.
  \label{eq:dfdt}
\end{align}
We now integrate over the transverse momentum components $p_x$ and $p_y$ 
and multiply by the number of spin states, $g_\chi = 2$, giving 
\begin{align}
  g_\chi \int \frac{dp_x dp_y}{(2\pi)^2} \mathbf{L}[f_\chi]
    &= g_\chi   \int\!\frac{dp_x dp_y}{(2\pi)^2} \frac{p_z^w}{E^w} \frac{\partial f_\chi}{\partial z^w}
     - g_\chi \biggl( \frac{\partial m}{\partial z^w} \biggr)
       \int\!\frac{dp_x dp_y}{(2\pi)^2} \frac{m(z^w)}{E^w}
                                        \frac{\partial f_\chi}{\partial p_z^w} \,.
\end{align}
Everything that we have done so far has been analytic and exact.  
We now introduce the ansatz from \cref{eq:ansatz}:
\begin{align}
  g_\chi \int \frac{dp_x dp_y}{(2\pi)^2} \mathbf{L}[f_\chi] 
    =&\, g_\chi \biggl( \frac{\partial }{\partial z^w} \FF(z^w,p^w_z) \biggr)
       \int \! \frac{dp_x dp_y}{(2\pi)^2}  \frac{p_z^w}{E^w}  f_\chi^\mathrm{eq}  \notag\\
    &+ g_\chi \FF(z^w,p^w_z) \int \! \frac{dp_x dp_y}{(2\pi)^2}
       \frac{p_z^w}{E^w} \biggl( \frac{\partial }{\partial z^w} f_\chi^\mathrm{eq} \biggr) \notag\\
    &- g_\chi \biggl( \frac{\partial m}{\partial z^w} \biggr) 
       \biggl( \frac{\partial }{\partial p_z^w} \FF(z^w,p^w_z) \biggr)
       \int \! \frac{dp_x dp_y}{(2\pi)^2}  \frac{m(z^w)}{E^w} f_\chi^\mathrm{eq} \notag\\
    &- g_\chi \biggl( \frac{\partial m}{\partial z^w} \biggr)  \FF(z^w,p^w_z)
       \int \! \frac{dp_x dp_y}{(2\pi)^2} \frac{m(z^w)}{E^w}
            \biggl( \frac{\partial }{\partial p_z^w} f_\chi^\mathrm{eq} \biggr) \,.
\end{align}
Using the Maxwell-Boltzmann approximation for $f_\chi$ then gives
\begin{align}
  g_\chi \int \frac{dp_x dp_y}{(2\pi)^2} \mathbf{L}[f_\chi] 
    \approx&\,  g_\chi \biggl( \frac{\partial }{\partial z^w} \FF(z^w,p^w_z) \biggr)
              \int \! \frac{dp_x dp_y}{(2\pi)^2}
                   \frac{p_z^w}{\sqrt{m^2 + p_x^2 + p_y^2 + (p_z^w)^2}} \, e^{-E^p/T} \notag\\
    &+        g_\chi \FF(z^w,p^w_z)
              \int \! \frac{dp_x dp_y}{(2\pi)^2}
                   \frac{p_z^w}{\sqrt{m^2 + p_x^2 + p_y^2 + (p_z^w)^2}} 
                   \biggl( \frac{\partial }{\partial z^w} e^{-E^p/T} \biggr) \notag\\
    &-        g_\chi \biggl( \frac{\partial m}{\partial z^w} \biggr) 
              \biggl( \frac{\partial }{\partial p_z^w} \FF(z^w,p^w_z) \biggr) 
              \int \! \frac{dp_x dp_y}{(2\pi)^2}
                   \frac{m}{\sqrt{m^2 + p_x^2 + p_y^2 + (p_z^w)^2}} e^{-E^p/T}  \notag\\
    &-        g_\chi \biggl( \frac{\partial m}{\partial z^w} \biggr)
              \FF(z^w,p^w_z)
              \int \! \frac{dp_x dp_y}{(2\pi)^2}
                   \frac{m}{\sqrt{m^2 + p_x^2 + p_y^2 + (p_z^w)^2}}
                   \biggl( \frac{\partial }{\partial p_z^w} e^{-E^p/T} \biggr) \notag\\
    =&\, \Bigg[ \Bigg( \frac{p_z^w}{m_\chi} \frac{\partial}{\partial z^w}
          - \bigg( \frac{\partial m_\chi}{\partial z^w} \bigg)
                                              \frac{\partial }{\partial p_z^w}  
          - \bigg( \frac{\partial m_\chi}{\partial z^w} \bigg) \frac{\vw}{T} \Bigg)
          \FF(z^w,p^w_z) \Bigg]
     \frac{g_\chi m_\chi T}{2\pi}
     e^{\big( \vw p_z^w - \sqrt{m_\chi^2 + (p_z^w)^2} \big) / T} \,.
  \label{eq:liouville}
\end{align}
Here we have used that $E^p \simeq E^w - \vw p_z^w$, which is reasonable 
for non-relativistic wall velocities.  
Since the final term in
\cref{eq:liouville} is proportional to the wall velocity, $\vw$, it is
important to keep the wall velocity in the Liouville operator, even when
dropping it elsewhere.

\subsection{The Collision Term}

We now turn to the collision term, $\mathbf{C}[f_\chi]$, which we evaluate in
the plasma frame, only transforming to the wall frame at the end.  We consider
the process $\chi(p^p) + \bar{\chi}(q^p) \to \phi(k^p) + \phi(l^p)$, where the
quantities in parentheses denote the momenta of the particles.  The collision
terms for the other processes, $\chi \phi \to \chi \phi$, $\chi \chi \to \chi
\chi$ and $\chi \bar{\chi} \to\chi \bar{\chi}$, can be derived by trivial
replacements.  Integrating over $p_x$ and $p_y$ and multiplying by the number
of spin states, $g_\chi = 2$, the collision term is 
\begin{align}
  g_\chi \int \! \frac{dp_x dp_y}{(2\pi)^2} \mathbf{C}[f_\chi]
    &= -\sum_\text{spins} \! \int \! \frac{dp_x dp_y}{(2\pi)^2} \,
         d\Pi_{q^p} \, d\Pi_{k^p} \, d\Pi_{l^p}
         \frac{(2\pi)^4}{2E_p^p} \delta^{(4)}(p^p+q^p-k^p-l^p) |\Mcal|^2 \notag\\
    &\qquad \cdot
       \Big[ f_{\chi_p}   f_{\bar{\chi}_q} (1\pm  f_{\phi_k}) (1\pm f_{\phi_l})
           - f_{\phi_k} f_{\phi_l} (1 \pm f_{\chi_p})   (1\pm f_{\bar{\chi}_q}) \Big] \,,
  \label{eq:collision}
\end{align}
where $\Mcal$ is the $CP$-invariant matrix element, and we have used  the
shorthand notation $E_p^p = [ (\pvec^p)^2 + m_\chi^2 ]^2$, $d\Pi_{q^p} \equiv
d^3q^p / [2 E_q^p \, (2\pi)^3]$, and $f_{\chi_p} \equiv f_\chi(t^p, \xvec^p,
\pvec^p)$, with $\chi_p \equiv \chi(p)$.  Analogous definitions are used for
the other momenta and distribution functions.

We neglect Pauli blocking and Bose enhancement for all species 
by setting $1 \pm f \approx 1$, 
and we assume that all species except for the initial DM particle $\chi(p)$ are
in equilibrium.  
For $\phi$, this is always true in the parameter region of interest to us;
for $\chi$, our numerical results show that $\chi$ does not deviate from equilibrium
by more than an $\mathcal{O}(1)$ factor, so the equilibrium approximation
is fairly accurate for any other $\chi$ particles in the process.
Since detailed balance holds for each momentum mode independently, 
$f_{\phi_k}^\mathrm{eq} f_{\phi_l}^\mathrm{eq} = f_{\chi_p}^\mathrm{eq}
f_{\bar{\chi}_q}^\mathrm{eq}$.  \Cref{eq:collision} thus simplifies to
\begin{align}
  g_\chi \int \! \frac{dp_x dp_y}{(2\pi)^2} \mathbf{C}[f_\chi]
    &= -\sum_\text{spins} \! \int \! \frac{dp_x dp_y}{(2\pi)^2}
         d\Pi_{q^p} \, d\Pi_{k^p} \, d\Pi_{l^p}
         \frac{(2\pi)^4}{2E_p^p} \delta^{(4)}(p^p+q^p-k^p-l^p)|\mathcal{M}|^2
         \Bigl[ f_{\chi_p} f_{\bar{\chi}_q}^\mathrm{eq}
              - f_{\chi_p}^\mathrm{eq} f_{\bar{\chi}_q}^\mathrm{eq} \Bigr] \,.
\end{align}
We can now integrate over $k$ and $l$ to obtain
\begin{align}
  g_\chi \int \! \frac{dp_x dp_y}{(2\pi)^2} \mathbf{C}[f_\chi]
    &= -g_\chi g_{\bar{\chi}} \int \! \frac{dp_x dp_y}{(2\pi)^2 2E_p^p} \, d\Pi_{q^p} \,
        4 F \sigma_{\chi\bar{\chi} \to \phi\phi}
        \Bigl[ f_{\chi_p} f_{\bar{\chi}_q}^\mathrm{eq} 
             - f_{\chi_p}^\mathrm{eq} f_{\bar{\chi}_q}^\mathrm{eq} \Bigr]    \notag\\
    &= -g_\chi g_{\bar{\chi}} \bigl[ \FF(z^w,p_z^w) - 1 \bigr]
       \int \! \frac{dp_x dp_y}{(2\pi)^2 2E_p^p} \, d\Pi_{q^p} \,
       4 F \sigma_{\chi\bar{\chi} \to \phi\phi} f_{\chi_p}^\mathrm{eq} f_{\bar{\chi}_q}^\mathrm{eq}
                          \label{eq:master-collision}  \,, \\
\intertext{where we have used the ansatz in \cref{eq:ansatz},
$\sigma_{\chi\bar{\chi} \to \phi\phi}$ is the relevant spin-averaged
cross-section, and where}
  F &= \frac{1}{2} \sqrt{(s - m_\chi^2 - m_{\bar{\chi}}^2)^2 - 4m_\chi^2m_{\bar{\chi}}^2} \,.
\end{align}
Although in principle we should replace $E^p$ in the equilibrium distribution functions 
with $\left(E^w - \vw p_z^w \right)$, 
the impact of $\vw$ is negligible in the collision term and we simply replace 
$E^p$ with $E^w$.
Making the Maxwell-Boltzmann approximation for $f^\mathrm{eq}$ 
then lets us perform the remaining integrals numerically reasonably quickly.

\section{Solving the Boltzmann equations}
\label{sec:solving-the-boltzmann}

We are now ready to solve the Boltzmann equation, \cref{eq:boltzmann},
with the Liouville operator on the left-hand side given by \cref{eq:liouville},
and the collision terms on the right-hand side given by
\cref{eq:master-collision} and similar terms for the other processes.
This equation is a partial differential equation (PDE) of the form
\begin{align}
    a(z^w, p^w_z) \frac{\partial \FF}{\partial z^w}
  + b(z^w, p^w_z) \frac{\partial \FF}{\partial p_z^w} &= c(\FF,z^w, p^w_z) \,.
\end{align}
PDEs of this form can be reduced to an infinite set of uncoupled 
ordinary differential equations (ODEs) using the method of characteristics. 
For any given starting point, a curve on the $(z^w, p_z^w)$ plane
can be defined via
\begin{align}
  \frac{d z^w(\lambda)}{d\lambda} = a(z^w, p^w_z) \,,
  \qquad
  \frac{d p_z^w(\lambda)}{d\lambda} = b(z^w, p^w_z) \,,
\end{align}
where $\lambda$ parameterises the curve.
The solution to the PDE along each curve can then be found by integrating the ODE 
\begin{align}
  \frac{d \FF\big(z^w(\lambda), p_z^w(\lambda)\big)}{d\lambda}
    &= c\big(\FF(\lambda), z^w(\lambda), p^w_z(\lambda)\big) \,.
\end{align}
To see this, note that
\begin{align}
  \frac{d \FF}{d\lambda}
    &= \frac{\partial \FF}{\partial z^w}   \frac{d z^w}{d\lambda}
     + \frac{\partial \FF}{\partial p_z^w} \frac{d p_z^w}{d\lambda}
     = a(z^w, p^w_z) \frac{\partial \FF}{\partial z^w}
     + b(z^w, p^w_z) \frac{\partial \FF}{\partial p_z^w}
     = c(\FF,z^w, p^w_z) \,.
\end{align}
Numerically, the full solution to the PDE on the plane can be found by
interpolating between solutions along several curves which span the region of
interest.  In Fig. 2 we show some of these curves.  Physically, in the
absence of collisions a particle with a given initial position and momentum,
$z^w$ and $p_z^w$, will travel along these curves in phase space as time
passes.  We see that particles starting outside the bubble and travelling
towards the bubble wall ($z^w \ll -\Lw$, $p_z^w > 0)$ are either reflected from
the bubble wall if $p_z^w \lesssim m_\chi^\mathrm{in}$, or penetrate the bubble
wall if $p_z^w \gtrsim m_\chi^\mathrm{in}$.   This is due to conservation of
energy.  Particles originating inside the bubble receive a boost of momentum as
they leave the bubble.

The boundary conditions for the PDE become initial conditions for the 
ODEs.  For particles outside the bubble which are approaching the bubble 
($z^w \ll -\Lw$, $p_z^w > 0$) we set $\FF = 1$, so the abundance is equal to the 
equilibrium abundance.  This assumes that $\chi$ are in equilibrium before the 
phase transition starts to take place.  
To fix the boundary condition for particles originating inside the bubble we 
assume that at $z^w \gg \Lw$ there is an identical parallel wall traveling 
in the opposite direction. 
To find the abundance at  ($z^w \gg \Lw$, $p_z^w < 0$),
we first compute the solutions along curves 
starting at ($z^w \ll -\Lw$, $p_z^w > m_\chi^\mathrm{in}$) and find the abundance 
deep inside the bubble, at ($z^w \gg \Lw$, $p_z^w > 0$). 
We then impose a periodic boundary condition at $z^w \gg \Lw$,
\begin{align}
  \FF(z^w \gg \Lw, p_z^w) &= \FF(z^w \gg \Lw, -p_z^w) \,. 
\end{align}
We neglect the effect of the wall velocity in this boundary condition, which is
a small modification.

\section{The fate of reflected dark matter}
\label{sec:reflected}

We have argued that only a tiny fraction of dark matter is able to pass through the ``filter'' and enter the broken-phase bubbles.  
The majority of dark matter is reflected back into the symmetric phase.  
What is the fate of this reflected dark matter?  

To ensure that dark mater particles do not accumulate in front of the advancing bubble walls, we require that they must annihilate away quickly via $\chi \bar{\chi} \to \phi \phi$.  
To simplify the estimates, we neglect $m_\chi$ and estimate the thermally-averaged cross section as $\langle \sigma v \rangle \approx (y_\chi^4 / 64 \pi T^2) \, \log (36T^2 / m_\phi^2)$.  
The thermally-averaged annihilation rate is calculated as $\Gamma \approx \langle \sigma v \rangle \, n_\chi^{\mathrm{out},\mathrm{eq}}$ where $n_\chi^{\mathrm{out},\mathrm{eq}} \approx 2 \times [3 \zeta(3) / 4 \pi^2] T_n^3$ is the equilibrium number density of (effectively massless) $\chi$ particles in front of the bubble wall.  
Requiring $\Gamma$ to be smaller than the Hubble expansion rate $H \sim \sqrt{g_\ast} T^2 / \Mpl$ implies a lower bound on the Yukawa coupling, which is approximately 
\begin{align}
	y_\chi \gtrsim \bigl( 8 \times 10^{-4} \bigr) \left( \frac{T_n}{1 \TeV} \right)^{1/4} \left( \frac{g_\ast}{106.75} \right)^{1/8} \left( \frac{\log 36 T_n^2 / m_\phi^2}{\log 36} \right)^{-1/4} \ . 
\end{align}
Since we are typically interested in $y_\chi = O(1)$, this condition is easily satisfied.  
Therefore we expect that the reflected dark matter does not accumulate in front of the bubble walls, but rather it annihilates away into $\phi$ particles.  
The $\phi$ particles maintain thermal equilibrium with the Standard Model plasma, and the entropy transfer from $\chi$ to $\phi$ heats the plasma.  
However, this is a negligible effect, since we consider phase transitions that are not strongly supercooled and there are roughly $g_{\ast S} \sim 100$ relativistic species at this time, while $\Delta g_{\ast S} \approx -4$ from the decoupling of $\chi$ and $\bar{\chi}$.  

\end{document}